\documentstyle[psfig]{article}
\begin{document}

\begin{center}
{\Large\bf Adaptive optics at the WHT}
\end{center}

\begin{center}
{\large C.R. Benn\footnote{Presented at the Sheffield 1999
conference `Instrumentation at the ING - the next decade'
(eds. N.A. Walton, S.J.S., Smartt, to be published in
New Astronomy Reviews)}}
\end{center}
\begin{center}
{\it Isaac Newton Group, Apartado 321, 38700 Santa Cruz de La Palma,
Spain}
\end{center}

\hrule

\large
{\bf Abstract}

The WHT is unusually well-placed for 
exploitation of adaptive-optics (AO) technology.  The site seeing is excellent
(median 0.7 arcsec), dome seeing is negligible, and preliminary studies
indicate that most of the atmospheric seeing originates in a well-defined 
layer at low altitude, which bodes well for future laser-guide-star 
AO.  The Durham group have built up extensive experience with 
natural-guide-star adaptive-optics experiments at the GHRIL Nasmyth focus, 
and the NAOMI common-user AO facility is due to be commissioned
at this focus early in 2000.  NAOMI will provide near-diffraction-limited 
imaging in the IR (Strehl $\sim$ 0.6, FWHM $\sim$ 0.15 
arcsec in K) and is expected to 
give significant correction in the optical (poorer Strehl, but similar FWHM).
NAOMI will perform better at short wavelengths than AO systems on other 
telescopes, and observers will require instrumentation that can exploit
this crucial advantage.

\vspace{5mm}

\vspace{3mm}
\hrule

\section{What is adaptive optics?}

`Active optics' and `Adaptive optics' 
improve image quality by correcting
distortions imposed on the wavefront during its passage
through the atmosphere and the telescope.
`Active optics' corrects for slow ($<\sim$ 0.01 Hz), low-order 
distortions, such as those caused by
sagging of optics when the telescope is tipped.  

{\it Adaptive} optics (AO) \cite{scam,beckers,jenkins}
improve seeing by
ironing out some of the wrinkling of the wavefront
caused by passage through turbulent layers in the atmosphere.
Typically, the wavefront is corrected by making independent tip, tilt 
and piston movements to each of $\sim$  10 - 100 elements 
of a deformable mirror in the collimated 
light beam, at $\sim$ 100 - 1000 Hz.
The corrections required are obtained by analysing
the wavefront from a bright star close on the sky 
to the object of interest, usually using a Shack-Hartmann wavefront sensor:

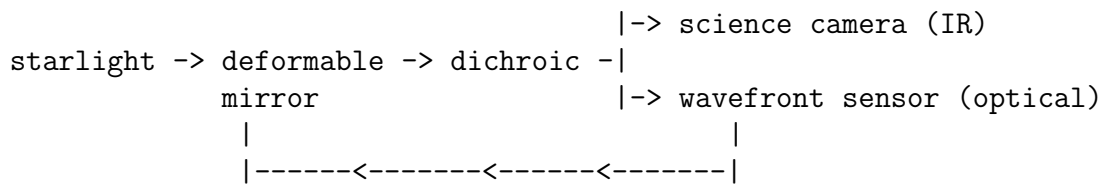
\begin{verbatim}
                                       |-> science camera (IR)
  starlight -> deformable -> dichroic -|
               mirror                  |-> wavefront sensor (optical)
                |                             |
                |------<-------<------<-------|
\end{verbatim}

The guide star must be bright enough for it to be detected in a fraction of a
second using light gathered by one  ($\sim r_0$-sized) sub-aperture,
i.e. $m_{GS} <\sim$ 14, with little dependence on telescope 
diameter, but depending strongly on the quality of correction required,
and the local atmospheric conditions.
The guide star must lie close enough to the target that the 
wavefronts from star and target suffer similar distortions.
This `isoplanatic radius',
typically $<$ 1 arcmin in the IR ($\propto \lambda^{6/5}$), 
determines sky coverage for a given AO system.

The quality of the seeing is parametrised by the Fried parameter $r_0$, 
$\approx$ the diameter of a telescope whose
diffraction-limited resolution is similar to the seeing.
$r_0$ = 20 cm yields optical 
seeing $\sim$ 0.5 arcsec.
$r_0 \propto \lambda^{6/5}$.
For a telescope with diameter $D < r_0$, diffraction dominates.
For $D \sim r_0$, image motion dominates 
(maximum gain for tip-tilt).
For $D > 4 r_0$, speckle dominates.

The quality of the correction is determined mainly by $r_0$, $m_{GS}$
and the bandwidth
of the AO system.
This quality is usually characterised by the
Strehl ratio of the resulting psf 
= (corrected peak intensity) /  
(unaberrated, i.e. diffraction-limited, peak intensity),
maximum value 1.0.
FWHM
can be a misleading measure, because partial correction 
(e.g. tip-tilt only) may yield an image with a narrow 
core superimposed on a diffuse plateau of emission, the latter including most
of the light.

With natural guide stars, 
and typical isoplanatic radii, sky coverage is $\sim$ 10\%
in K and $\sim$ 0.5\% in V. 
The exact numbers depend strongly on atmospheric conditions and acceptable
Strehl, and on galactic latitude (beware of values quoted out of context!).
An artificial guide star, created by laser illumination of the mesospheric
Na layer $\sim$ 100 km up, allows observing anywhere on the sky.
A laser guide star cannot be used to measure the tip-tilt correction
(because of the symmetry of the path up and down), so a natural
guide star is still needed for this.  However, the radius over which
guide stars can be used for tip-tilt, the isokinetic radius, is
$>>$ the isoplanatic radius.
No astronomical observatory yet has an effective laser-guide-star system,
although several are at the prototype stage.  

Compared with HST, ground-based AO can deliver better resolution 
and larger collecting area, but the dynamic range is poorer.

\section{Science drivers}
The potential of an advance in astronomical technology is often
judged by how much it enlarges the observational parameter space.
The range of measured optical 
angular diameters $\theta$ (Fig. 1) is small compared to say, the range
of wavelengths used by astronomers, or the range of measured optical
intensities (both $>$ 10 decades).
Thus, 
in terms of expanding the observational parameter space, an improvement
of a decade in angular resolution (1 arcsec $\rightarrow$ $\sim$ 0.1 arcsec) 
is substantial.
Fig. 1 also suggests that 
an AO system, like a large new telescope, 
will find application in many different areas,
and this is reflected in the comparable amounts of space devoted
in AO science cases to 
solar-system, stellar and extragalactic topics.

Several AO systems are now up and running.  What are they being used for?
In the years 1993, 94, 95, 96, 97 and 98, the numbers of AO-science
papers published were
3, 2, 5, 8, 24 and 19 \cite{keck}, indicating a fast-growing field. 
The titles of the 19 AO-science publications in 1998 
cover, as expected, an eclectic range of topics:
comets, asteroids, HII regions, circumstellar disks,
dwarf galaxies, AGN and starburst galaxies, QSO hosts and
gravitational lenses.

The use of AO to study a wide range of physical phenomena suggests
that AO systems will require general-purpose science cameras i.e.
both imaging devices and spectrographs operating over the full range of
wavelengths at which there is appreciable correction.

\begin{figure}
\centering
\psfig{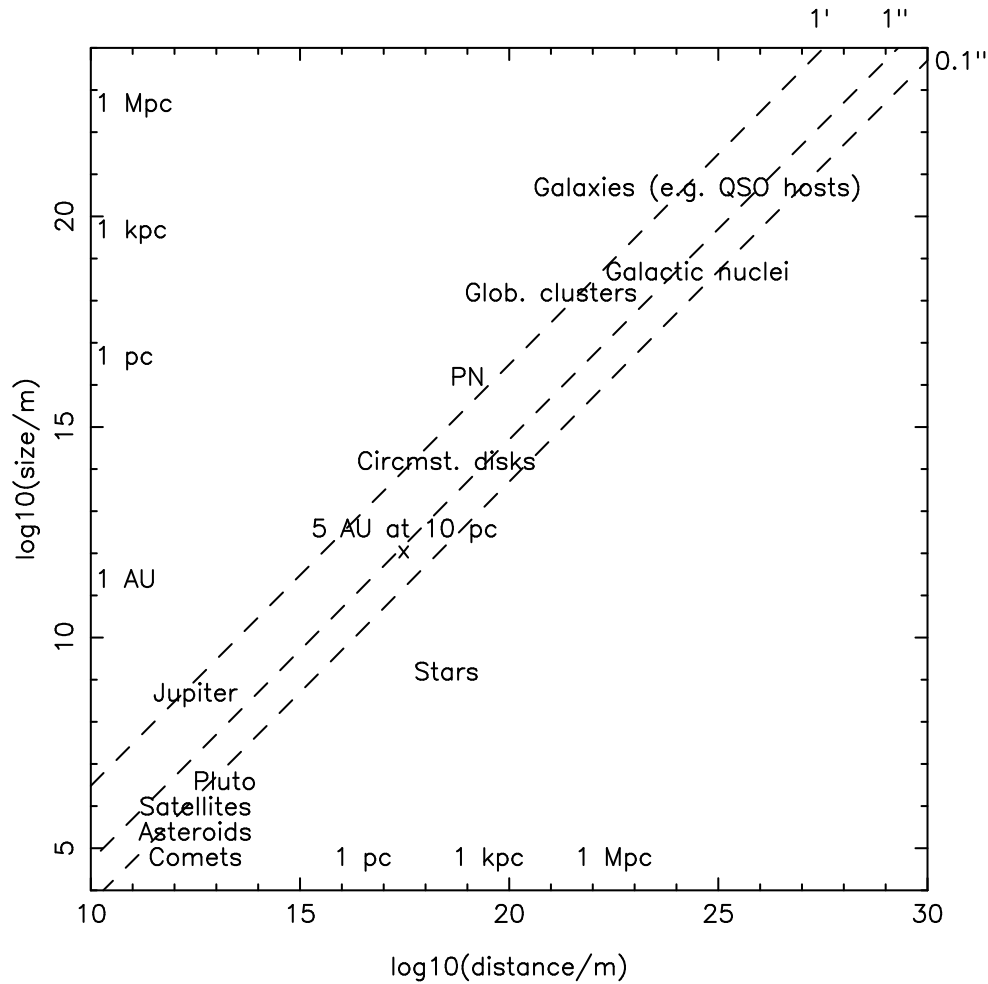}
\caption{Measured angular sizes of astronomical objects span a small
range in the optical.  Gravitational stability imposes the requirement that
astronomical objects are generally much smaller than their separations i.e.
the nearest other planets, stars and galaxies typically subtend
$\theta <<$ 1 radian, while the earth's atmosphere limits measured
$\theta >\sim$ 1 arcsec.
The dashed lines indicating angular size do not take into
account cosmological distortions at large scales.}
\end{figure}

\section{What are other observatories doing?}
AO systems are being built for most of the $\approx$ 29 operating
or planned optical telescopes with primary-mirror diameter $>$ 3.5 m.
The Calar Alto 3.5-m, CFHT, ESO 3.6-m and Keck II all have functioning
AO systems.

CFHT's AO system, PUEO, has been operating successfully since 1996,
and comprises a 19-element deformable mirror operating at 1 kHz
(suited to the weak, fast turbulence over Mauna Kea), feeding an IR
camera and an integral-field spectrograph.
Recently, AO correction has been extended to the optical, 
achieving 0.24 arcsec FWHM in V (Strehl 0.1) \cite{cfht}.
This shows that correction in the 
optical is achievable, and bodes well for NAOMI, with its design emphasis
on operation at short wavelengths.

Calar Alto's ALFA 
has also been in operation since 1996, and delivers
high Strehl (0.6) in K band.  Spectroscopy 
of two objects separated by 0.26 arcsec
was recently reported \cite{alfa}.
Calar Alto is also well advanced with a laser-guide-star system.
The main driver for AO at Calar Alto is alleviation of the poor site seeing, 
usually $>$ 1 arcsec.

Keck II's AO system has just been commissioned (Feb 1999) and delivered
0.04 arcsec FWHM in K band (Strehl 0.3).  
The deformable mirror has 349 actuators.
A similar system (currently in Waimea) exists for Keck I.
One of the drivers for AO at Keck is the need to maximise
wavefront correction before using the light for interferometry
between Keck I and II.

\section{Quality of the La Palma site}
The quality of the seeing at the WHT has been measured as part of
ING's `Half-arcsec programme', under which causes of dome seeing
have been identified and minimised.
A Shack-Hartmann camera (JOSE) was used at the
Nasmyth focus to measure the distribution of $r_0$
\cite{wilson}, and it was found to be
almost indistiguishable from distributions measured at CFHT 
and in Chile, median seeing 0.7 arcsec.
The distribution of $r_0$ at the WHT is also identical to that
measured 
by a DIMM seeing monitor outside the WHT, 
implying that dome seeing is negligible.

The vertical structure of the turbulence is not well known at any
site, but the few measurements available for La Palma (JOSE, see also
\cite{munoz}) suggest that
the turbulence is dominated by low-lying layers, which bodes well
for correction using laser-guide-stars.

\section{NAOMI - adaptive optics at the WHT }
NAOMI, the WHT's common-user AO system \cite{wht}, 
is scheduled for commissioning early 2000.
It's predecessor ELECTRA has had several successful commissioning runs at the 
WHT, achieving FWHM 0.15 arcsec in J, H and K.

NAOMI is designed to deliver near-diffraction-limited performance at
a wavelength
of 2 $\mu$ using natural guide stars.
Specifically, it should deliver, under median-seeing conditions,
Strehl $>$ 0.25 over 50\% of the sky, and Strehl
$>$ 0.7 over 5\% of the sky.
At shorter wavelengths, the performance will not be
diffraction limited, but there will be partial correction,
probably FWHM $>$ 0.2 arcsec at 0.7 $\mu$.
The isoplanatic radius at K (where the Strehl ratio falls to half its on-axis
value), will be $\sim$ 20 arcsec.

NAOMI is expected to perform well in the optical, compared to other
AO systems, thanks to 
the (76-element) 
segmented deformable mirror (most AO systems use a mirror with
continous face-sheet) and to the unusually accurate positioning of the
mirror segments.  
NAOMI will operate at the Nasmyth focus and will
have separate ports for optical and IR cameras.
Initially, imaging will be available at the IR port 
using ING's new IR camera INGRID (0.04 arcsec/pixel, field 40 arcsec),
and may be available
at the optical port using a test camera (field 15 arcsec).  
Optical spectroscopy will be available via the TEIFU fibre feed to WYFFOS,
(field 6 arcsec, at 0.25 arcsec/pixel, or 3 arcsec at 0.13 arcsec/pixel).
TEIFU was
partly commissioned June 1999.
No IR spectrograph is currently planned, but a number of 
inexpensive options are
being considered, including the possibility of
IR spectroscopy with WYFFOS, or with a warm grating in front of INGRID.

\section{What next?}
With natural guide stars, NAOMI will offer sky coverage of only
a few \% in the IR, and $<$ 1\% in the optical, and this coverage is
biased to low galactic latitudes.  
Some improvement in coverage is possible through
enhancements of the existing
system e.g. by conjugation to image the dominant turbulent layer
at the deformable mirror, or
by deployment of a wavefront-sensor detector with very low readout noise
(allowing use of fainter guide stars).
Laser guide stars (LGS), however, offer the best chance of obtaining
reasonable sky coverage in the optical.
A team from IC/Durham will use a (weak) LGS in autumn 1999 to
characterise the sodium layer above La Palma.  LGS AO 
could be available at the WHT sometime after 2002.
LGS AO in the optical will be much more to difficult to achieve on an 8-m
telescope because of the severe shrinking of the isoplanatic patch caused
by laser and starlight traversing different paths through the atmosphere
(`cone effect').
Chris Dainty and Andy Longmore discuss LGS AO
in their presentations (this volume).

\end{document}